# Direct observation of electron emission as a result of a VVV Auger transition in the valence band of Graphene

V. A. Chirayath[1,*], V. Callewaert[2], M. D. Chrysler[1], A. J. Fairchild[1], R. W. Gladen[1], A. D. Mcdonald[1], S. K. Imam[1], K. Shastry[1], A. R. Koymen[1], R. Saniz[2], B. Barbiellini[3], K. Rajeshwar[4], B. Partoens[2] and A. H. Weiss[1]

[1]Department of Physics, University of Texas at Arlington, Arlington, Texas – 76019, USA
[2]Department of Physics, Universiteit Antwerpen, Antwerpen 2020, Belgium
[3]Department of Physics, Northeastern University, Boston, Massachusetts 02115, USA
[4]Department of Chemistry and Biochemistry, University of Texas at Arlington, Arlington, Texas – 76019, USA

**We report the first direct observation of electron emission into the vacuum as a result of a VVV Auger transition resulting from the relaxation of a deep hole in the valence band. A beam of low energy (<1.25eV) positrons was used to deposit positrons onto the surface of samples consisting of single layer graphene, multi-layer graphene and graphite. The distribution of electrons emitted from the samples as a result of the annihilation of the positron showed peak extending up to ~12 eV with a maximum at ~4eV. The observed peak was ~17 times larger than the previously observed annihilation induced C KVV peak. An analysis based upon a density functional theory calculation of the positron annihilation rates indicates that the width and intensity of the peak is consistent with electron emission resulting from VVV Auger transition excited by the annihilation of valence band electrons. Good agreement was found between the data from the single layer graphene on Cu surface with a theoretical line shape found from a self-folding of the density of states for a free standing graphene layer. The agreement between the theoretical and measured intensities for the KVV and VVV transitions indicates that the branching ratio for holes to decay via an Auger transition is nearly the same in both cases (i.e. close to 100%). Our results suggest the possibility of using annihilation induced VVV Auger spectroscopy to study the properties of the local density of states and the hole decay processes in materials in which the valence band width exceeds the work function.**

Auger transitions are known to be the predominant mode of relaxation for atoms containing holes in core levels with binding energies in the 30eV to 2000eV range [1]. Similarly, Auger transition are thought to play a critical role in the recombination of excitonic states in the range of 1 -1000 meV [2-6]. Here we present experimental results showing electron emission into the vacuum from the surface of single layer graphene and graphite as a result of Auger transitions starting with the filling of deep holes in the valence band (VB) by another electron higher up in the VB. While it has been posited that such processes (termed VVV Auger transitions) might contribute to the low energy part of photon or electron induced electron spectra, it has, up to now not been possible to directly separate out and measure the intensity of these contributions because of the low energy of transitions involved and large background resulting from primary beam induced secondary electrons in this energy range [7]. These measurements were made possible by the elimination of almost all background by creating the initial valence hole through an annihilation process with a positron injected into a surface state using a positron beam whose energy (~1.25 eV) was below the threshold for impact induced secondary electron emission [8]. The annihilation induced hole is filled by a second electron in the band and results in the excitation of an electron to an unoccupied state as shown in the schematic for a VVV Auger process in Fig.1.

Neglecting correlation effects, the kinetic energy, $KE_A$, of electrons emitted into the vacuum via a VVV Auger transition can be determined from energy conservation to be:

$$KE_A = E_1-E_2-E_3-\phi \qquad (1)$$

where $E_1$, $E_2$, and $E_3$ are the binding energies of the electronic states involved in the transition (with reference to the Fermi level) and $\phi$ is the work function. Since all three of the energy levels are in the VB, not all values for $E_1$, $E_2$, $E_3$ can fulfil the requirements for an Auger transition that can result in an electron with sufficient energy to overcome the work function and be emitted into the vacuum. From equation (1) it can be seen that the maximum kinetic energy for such an electron emitted into the vacuum is $KE_{Auger-max} = W-\phi$ where W is the width of the valence band and that electrons can be emitted into the vacuum via a VVV transition for any material for which $(W-\phi)>0$. In the case of graphene, the width of the valence band (~20eV) and the value of the work function ($\phi \cong 4.5$ eV), make it possible for electron generated by a VVV process to have a maximum kinetic energy of ~15 eV. Our calculations, described in more detail below, indicate that a significant fraction (~25%) of electrons generated via energetically accessible VVV Auger transitions have sufficient energy to escape into the vacuum.

The measurements were performed using the UT Arlington Time of Flight Positron Annihilation Induced Auger Spectrometer, consisting of a low energy positron beam equipped with TOF spectrometer with a ~1m flight path. An axial magnetic field of ~0.0040 Tesla is used to confine both the incident positrons and outgoing electrons to paths along the beam axis resulting in efficient transport. A permanent magnet is used to create a 0.045 T field at the sample. The gradient in the field redirects the velocity of the outgoing electrons along the axis of the TOF spectrometer while allowing for the collection and detection of electrons emitted into ~ $2\pi$ sr. The system was configured specially so as permit both the transport of a very low energy positron beam to the sample and the measurement of the energy of electrons emitted from the sample over a wide range of energies (0.1eV – 800eV). The apparatus and low energy settings used have been described in detail in a previous publication [8]. The TOF spectra were converted to energy using an energy conversion function containing parameters that are adjusted to fit an experimentally derived calibration curve. All of the spectra shown in this paper have been normalized by dividing by a number proportional to the number of detected annihilation gamma rays from the sample.

Figure 2 shows the TOF spectra of positron annihilation induced Auger electrons emitted from a sample consisting of 1 layer of graphene on a Cu substrate and the same sample after sputtering to remove the graphene and expose a clean Cu surface. The TOF spectrum from polycrystalline Cu has three main features; the Auger peaks corresponding to $M_{2,3}VV$ and $M_1VV$ transitions at 60 eV and 108 eV respectively and the low energy tail (LET). Spectral intensity in the LET has contributions from the inelastic loss of the higher energy Auger peaks. The spectral weight under the inelastic tail (defined up to ~12 eV) is 2.4 times the intensity in $M_{2,3}VV$ peak at 60 eV. Auger peaks corresponding to common impurities, carbon or oxygen were not detected in the PAES spectra which shows that the polycrystalline Cu surface to be clean.

The spectra of single layer graphene show Auger peak corresponding to KVV transition in Carbon at 263 eV. It also shows the presence of adsorbed oxygen on the surface through the Auger peak corresponding to the KVV transition (503 eV) in oxygen. The data also shows a peak corresponding to the $M_{2,3}VV$ Auger transition in Cu which has ~ 0.11 times the intensity under the $M_{2,3}VV$ peak observed for the clean Cu surface. The intensity under the $M_{2,3}VV$ peak was calculated after subtracting the inelastic contribution of higher energy Auger electrons (in this case from carbon and oxygen) towards the energy region corresponding to $M_{2,3}VV$ peak. The presence of the Cu Auger peak indicates that there is some overlap of the positron surface state wave function with Cu atoms even though the graphene film was found to be continuous using AFM measurements across various regions of the sample.

The broad peak at ~1.65 µs in the TOF spectra of graphene corresponding to an energy ~ 4 eV in the energy spectra is the low energy peak due to the VVV Auger process. Note that this peak is absent in the spectra of clean Cu and that this low energy peak cuts off by ~ 12 eV. The integrated intensity observed from the graphene samples in the 0eV - 12 eV range show more than an order of magnitude larger than the integrated intensity of the most prominent higher energy peaks (C KVV, O KVV). Consequently, the LET from these higher energy peaks cannot account the large intensity observed in the low energy peak in graphene. A similar broad feature was also seen for highly oriented pyrolytic graphite [9] which rules out any significant contribution from the copper substrate to the low energy peak seen in graphene. LET arising out of inelastic loss processes and cascade processes of higher energy Auger peaks was subtracted from the low energy VVV Auger peak by fitting the LET of the Cu PAES spectrum to a model function and scaling it to the spectral intensity under energy regions from 30 eV to 750 eV (which encompasses the Auger peaks of C, O and Cu). In order to quantify the efficiency of the VVV Auger process, the efficiency of the spectrometer that includes the efficiency of the detectors and efficiency of transport need to be considered along with the absolute number of positrons annihilating with valence electrons in graphene.

The TOF spectrometer used in the experiment was simulated using SIMION 8.1 ® [10] to explore the energy dependence of the electron transport efficiency of the spectrometer. A drop in transport efficiency was found for electron energies above 500 eV and for energies below 1 eV. Corrections were made to the energy spectra to take this loss of efficiency into account before background subtraction. The number of positrons annihilating with graphene valence electrons depends on the number of positrons which reach the sample, the fraction of positrons which forms positronium on graphene and the fraction which annihilates with core electrons. These quantities can however be avoided by taking the ratio of the intensity under the VVV Auger with the intensity under the C KVV Auger transition. By taking the ratio, many parameters which are not dependent on the energy of the electron gets cancelled out making the quantification detector independent. The comparison to C KVV is inspired by the fact that almost every core hole created in K shell of C decay through an Auger process. This number can be compared directly to a theoretically calculated quantity like the ratio of the positron annihilation rate with electrons belonging to different orbitals in graphene as the intensity ratio will be proportional to the ratio of annihilation rates as well as to the efficiency of the VVV Auger process. From the knowledge of positron annihilation rates it will be possible to deduce the efficiency of the VVV Auger process in graphene.

We have compared our experimental results with first-principles calculations of the line shape of the annihilation induced VVV transition and the relative intensities of the VVV and CVV peak intensities which will be described in detail elsewhere. This comparison provided confirmation of the annihilation induced VVV Auger emission mechanism and yielded an estimate of the Auger branching ratio for the initial valence hole from the relative intensities of the annihilation induced CVV and VVV Auger peaks. The calculations can be decomposed into 3 steps. In the first step, the positron surface state wave function is calculated. In the second step, the annihilation rates of the surface state positrons with the valence and core levels are computed through the evaluation of wave function overlap integrals. In the third step, the-hole distribution deduced from the relative annihilation rates is used as a starting point in the calculation of the line shape of the annihilation induced VVV transition applying a formalism similar to that used for Auger neutralization [12].

The calculations of the surface state wave functions and the annihilation rates, $\lambda$, with core and valence electrons were performed within the framework of the two-component electron-positron density functional theory, using a non-local weighted density approximation to describe the electron-positron correlation effects [13] along with Drummond enhancement of the annihilation rate [14]. The distribution of initial, annihilation induced, holes is determined by the partial annihilation rate $\lambda(\varepsilon_h)$ of the positron with the electronic states at a given energy $\varepsilon_h$. It was assumed that the one-hole initial state created by the annihilation process decays to a two-hole final state through an Auger process whose spectra was assumed to be described by a self-convolution of the one-particle density of states [15, 16]. This type of convolution integral has been successful in describing band-like CVV Auger spectra. Hole-hole interaction as described by Cini and Sawatzky [17] in the final two-hole state was not considered.

For a clean Cu surface, we find that the positron is located in a surface state and has a non-negligible overlap with only the first two atomic layers of the Cu (111) surface. For single layer graphene on a Cu substrate, the calculations show that the positron state is located at the vacuum side of the sample, with a small but non-negligible overlap with states of the Cu (see Fig. 3(a)) corresponding to 92% of the annihilations occurring with electrons from the graphene over-layer and 8% with the Cu substrate. The density of states (calculated for a free standing single layer of graphene) used in the convolution integral is shown Fig. 3(b) along with the distribution of annihilation induced initial hole states determined from a calculation of the state dependent annihilation rates.

A comparison of the measured and theoretical VVV Auger line shape from a single layer of graphene on Cu is shown in Fig. 4. The inelastic tail from high energy Auger peaks has been subtracted from the spectral intensity below 12 eV in the experimental curve. The calculated Auger spectrum has been used as an input to a charged particle trajectory simulation of the experimental setup to take into account instrumental broadening. The output of the simulation is a TOF spectrum which has been converted to energy spectrum and smoothed using the same algorithm employed on the experimental TOF spectrum.

There is an excellent match in the overall width and the shape of the spectrum with some deviation apparent at low energies where the measured energy distribution exceeds the theoretical. The best agreement was found by adding a small contribution from what is assumed to be electrons scattered as they exit the surface. This was done by calculating a Shirley function (shown in red in Fig. 4) from the measured energy distribution and adding it to the calculated annihilation induced VVV line shape. It is likely that extending the theory further to include consideration of the hole-hole interaction in the two hole final state through a Cini - Sawatzky formalism [17], introducing the effect of oxygen adsorption on graphene, comparing the actual electron work function of the sample with the calculated value and better modelling of the inelastic scattering would lead to better agreement between the theory and experiment at lower energies.

The possible effect of final state hole-hole interactions may be seen in Fig. 5 which shows the measured energy spectrum of annihilation induced low energy electrons from single layer graphene on Cu, 6-8 layers of graphene on Cu and HOPG (graphene on graphite). The observed shift in the spectrum to lower energies in going from single layer graphene on Cu to multilayer graphene on Cu to graphene on graphite is consistent with an expected shift in the spectrum of electrons emitted through a VVV Auger transition as a result of the an increase in the final state hole-hole interactions as metallic screening is decreased [18].

Table 1 lists the measured and theoretically derived values of the ratios of the integrated intensities of electrons emitted through the C VVV Auger transition to integrated intensity emitted through the C KVV Auger transitions for a single layer of graphene on Cu. Both the C VVV and C KVV intensities were calculated assuming that all holes that could decay via an Auger transition that could result in an electron emitted into the vacuum did decay via the Auger process. Previous calculations performed by McGuire [11] found that almost all (>99%) holes in the K shell of C decay via a KVV Auger process. Thus the excellent agreement between the measured and calculated ratios of C VVV to C KVV intensities provides strong evidence that our assumption that almost all holes in the valence band energetically capable of producing Auger electrons emitted into the vacuum (i.e. holes deeper than 4.5 eV below the Fermi level) decay via Auger transitions is justified. If a significant fraction of these deeper holes in valence band decayed via a competing process, than the measured ratio of C VVV to C KVV would be significantly lower than the theoretical ratio. For example if a deep hole decayed primarily via a "bubbling up" process consisting of a large number of low energy Auger excitations of electron hole-pairs near the Fermi level then the emission of electrons via VVV Auger process would be suppressed. It may be that filling deep holes via multiple low energy Auger processes is precluded in graphene due to the reduced density of states at the Fermi level. It would be of great interest to study the relative annihilation induced VVV intensities from different materials to see if the high efficiency of the Auger process seen in graphene is a universal property.

In this paper we have presented the first direct measurement of electrons emitted as a result of (VVV) Auger transitions taking place wholly in the valence band of graphene and graphite. As mentioned above, the ability to use a positron beam energy below the secondary electron emission threshold played a key part in allowing us to eliminate the beam induced secondary electron background and to separate out and directly measure the contributions of VVV Auger transition. Another important advantage of using very low energy positrons is that almost all positrons implanted at low energies become trapped in an image potential well localized just outside the surface prior to annihilation. This results in almost all annihilation induced Auger electrons originating in the top most layer elimination almost all spectral contributions from extrinsic loss processes associated with Auger electrons exciting from below the surface layer. The surface sensitivity of PAES was particularly helpful in the measurement of single layer graphene in which case the trapping of positrons at the surface minimized spectral contributions from the Cu substrate. We note that, due to the relatively small interaction cross section of 511 keV gamma rays, contributions from gamma induced secondary electrons can be neglected.

The measurement showed a low energy peak in C carbon peaking at ~3 eV and extending up to ~12 eV that was ~17 times larger than the positron annihilation induced Auger C KVV peak. A theoretical energy distribution calculated using a model for the VVV Auger transition which takes into account the density of the annihilation induced hole states and the density of filled and empty electron states results in model spectra that is in good agreement with our experimental results suggesting that the VVV Auger transition in graphene is band-like. A detailed calculation showed that this ratio was consistent with the calculated ratio of the positron annihilation rates with K shell versus Valence Band electrons in C with the assumption that all of the VB holes created through annihilation decayed via Auger processes. Our results present strong evidence that the Auger process is the dominant mode of relaxation for VB holes whose binding energy is in a range that could result in the emission of an electron into the vacuum through an Auger process. These results suggest that deep hole lifetimes in graphene will be short and that carrier multiplication via Auger processes will be highly efficient. Since VVV process resulting in the emission of electrons can occur whenever the valence band width exceeds the work function it should be possible to apply our technique to the measurement of VVV transitions in a significant number of materials including Si. This opens up the possibility of using annihilation induced VVV Auger spectroscopy to study the properties of the local density of states and the decay process of holes in the valence band in these materials.

## Acknowledgment

This work was supported by the grant NSF DMR 1508719. A.H.W and A.R.K gratefully acknowledge support for the building of advanced positron beam through the grant NSF DMR MRI 1338130.

## References

1. Briggs, D. & Riviere, J. C. Spectral Interpretation. In Practical Surface Analysis Edn. 2 Vol. 1 (eds. Briggs, D. & Seah, M. P.) (John Wiley and Sons, Sussex, England, 1990).
2. Iveland, J., Martinelli, L., Peretti, J., Speck, J. S., & Weisbuch, C. Direct Measurement of Auger Electrons Emitted from a Semiconductor Light-Emitting Diode under Electrical Injection: Identification of the Dominant Mechanism for Efficiency Droop *PRL* 110, 177406 (2013). http://dx.doi.org/10.1103/PhysRevLett.110.177406
3. Plötzing, T. et al. Experimental Verification of Carrier Multiplication in Graphene *Nano Lett.* 14, 5371–5375, (2014). http://dx.doi.org/10.1021/nl502114w
4. Winzer, T. & Mali´c, E. Impact of Auger processes on carrier dynamics in graphene *Phys. Rev. B* 85, 241404(R) (2012). http://dx.doi.org/10.1103/PhysRevB.85.241404

5. Tielrooij, K. J., et al. Photoexcitation cascade and multiple hot-carrier generation in graphene *Nature Physics* 9, 248–252 (2013). http://dx.doi.org/10.1038/nphys2564
6. Brida, D., et al. Ultrafast collinear scattering and carrier multiplication in graphene *Nature Communications* 4, 1987 (2013). http://dx.doi.org/10.1038/ncomms2987
7. Braicovich, L. Comparing valence Auger line shapes and photo emission spectra: The Cd and CdO cases. In Emission and scattering techniques: Studies of inorganic molecules, solids and surfaces (eds. Day, P.) (D. Reidel Publishing company, Dordrecht, Holland, 1980).
8. Mukherjee, S., et al. Time of flight spectrometer for background-free positron annihilation induced Auger electron spectroscopy *Review of Scientific Instruments* **87**, 035114 (2016). http://dx.doi.org/10.1063/1.4943858
9. Gladen, R. W., et al Time of flight spectra of electrons emitted from graphite after positron annihilation *14th International Workshop on Slow Positron Beam Techniques and Applications (SLOPOS14), Japan* Fujinami M et al *Journal of Physics. Conference Series*. (2017).
10. Manura, D. & Dahl, D. SIMION (R) 8.0 User Manual (Scientific Instrument Services, Inc. Ringoes, NJ 08551, <http://simion.com/>, January 2008).
11. McGuire, E. J. K-Shell Auger Transition Rates and Fluorescence Yields for Elements Be-Ar, Physical Review 185, 1 (1969). https://dx.doi.org/10.1103/PhysRev.185.1
12. Hagstrum, H. D., et al. Theory of Auger Neutralization of Ions at the Surface of a Diamond-Type Semiconductor *Phys. Rev.*, **122**, 83 (1961). https://dx.doi.org/10.1103/PhysRev.122.83
13. Rubaszek, A. Electron-positron enhancement factors at a metal surface: Aluminum *Phys. Rev. B* **44**, 10857 (1991). https://doi.org/10.1103/PhysRevB.44.10857
14. Drummond, N. D., Lo´pez R1´os, P., Needs, R. J., and Pickard, C. J. Quantum Monte *Carlo Study* of a Positron in an Electron Gas *PRL* **107**, 207402 (2011). https://doi.org/10.1103/PhysRevLett.107.207402
15. Lander, J. Auger Peaks in the Energy *Sp***ctr**a of Secondary Electrons from Various Materials *Phys. Rev.*, **91**, 1382 (1953). https://doi.org/10.1103/PhysRev.91.1382
16. Weiss, A. H., Sundaramoorthy, R., Hulbert, S. L., Bartynski, R. A., Modeling of the energy spectra of individual steps of the $L_{23} \rightarrow M_{2,3}M_{2,3} \rightarrow M_{2,3}VV \rightarrow VVVV$ cascade chain in MnO *J. Electron Spectrosc. Relat. Phenom.* **161**, 160 (2007).
17. Cini,, M. Density of states of two interacting holes in a solid *Solid State Commun.* **20**, 605 (1976). http://dx.doi.org/10.1016/0038-1098(76)91070-X. Sawatzky, G. A. Quasiatomic Auger spectra in narrow band materials *Phys. Rev. Lett*. **39**, 504 (1977). https://doi.org/10.1103/PhysRevLett.39.504
18. Ramaker, D. E. The past, present, and future of auger line shape analysis *Critical Reviews in Solid State and Materials Sciences* **17,** 211-276 (1991). http://dx.doi.org/10.1080/10408439108243752

Table 1: Comparison of the experimentally and theoretically calculated ratio of intensity of the C VVV Auger peak to the C KVV Auger peak in single layer graphene .

| Intensity ratio | Experiment | Theory |
|---|---|---|
| $\dfrac{C\ VVV}{C\ KVV}$ | 17 | 17 |

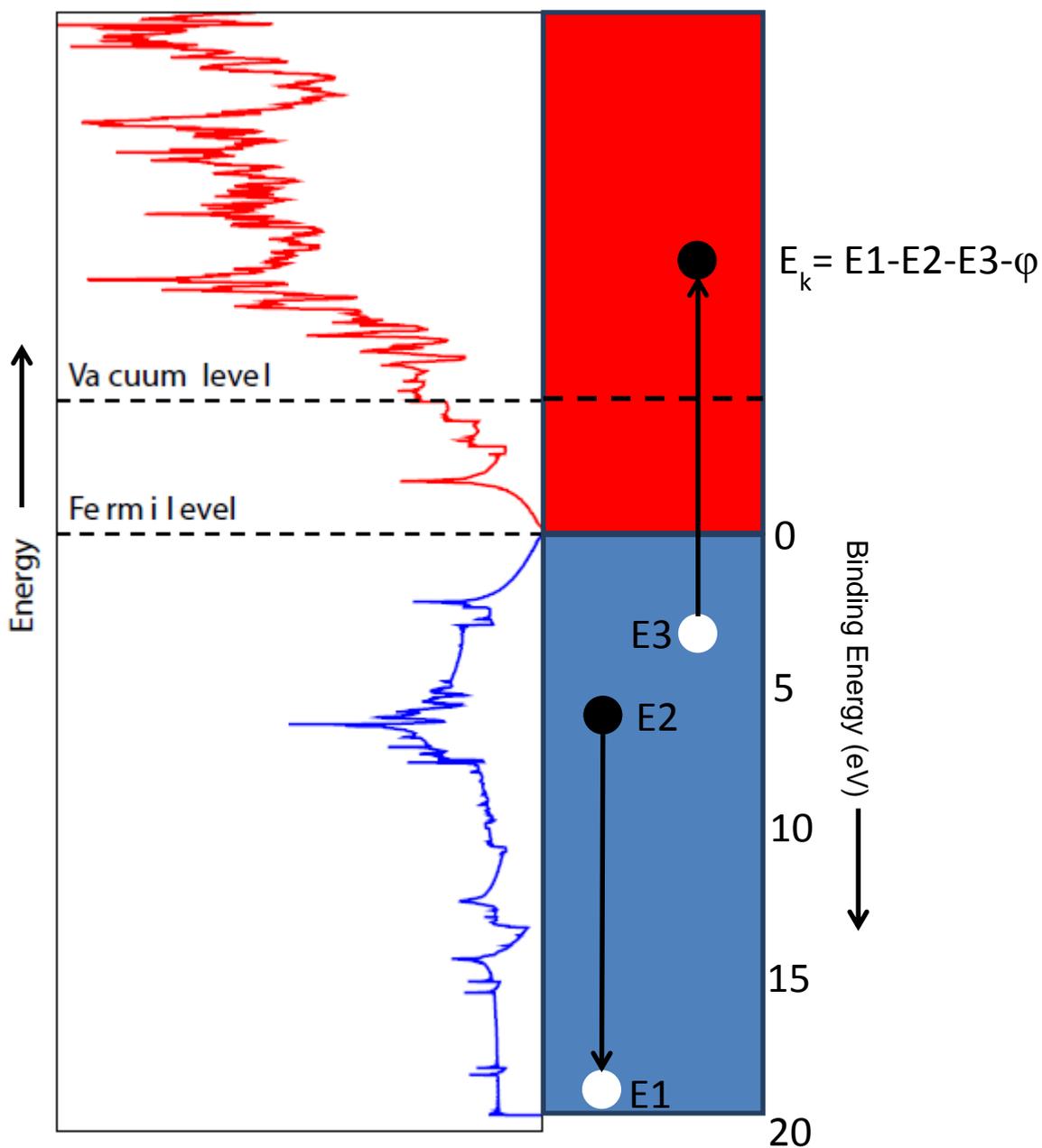

Fig.1. Schematic representation of VVV Auger transition is shown here. A deep hole in the valence band gets filled by an electron from higher in the valence band. The energy associated with this transition is coupled to a third electron in the valence band which then escapes from the surface of graphene with an energy equal to $E_1-E_2-E_3-\varphi$ where $E_1$, $E_2$ and $E_3$ are the binding energies of the electrons involved in the transition (referenced to the Fermi level) and $\varphi$ is the electron work function of the single layer graphene. (color online)

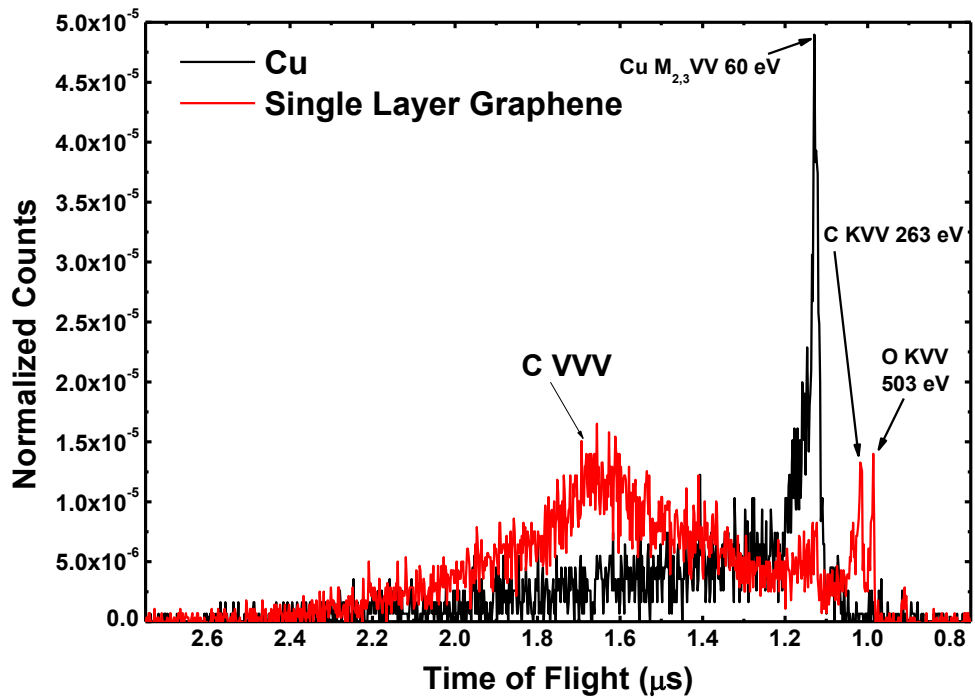

Fig.2 TOF spectra of electrons emitted from single layer graphene (on polycrystalline Cu substrate) and clean polycrystalline Cu after positron annihilation. The broad peak at 1.65 μs (~4eV) corresponds to a VVV Auger transition initiated by the annihilation of electron deep in the valence band of graphene by a positron trapped at the surface. The positrons were incident on the sample with energies < 1.25 eV. Peaks in the time of flight spectrum corresponding to annihilation induced Auger transitions from Cu, C and Oxygen are identified in the plot.

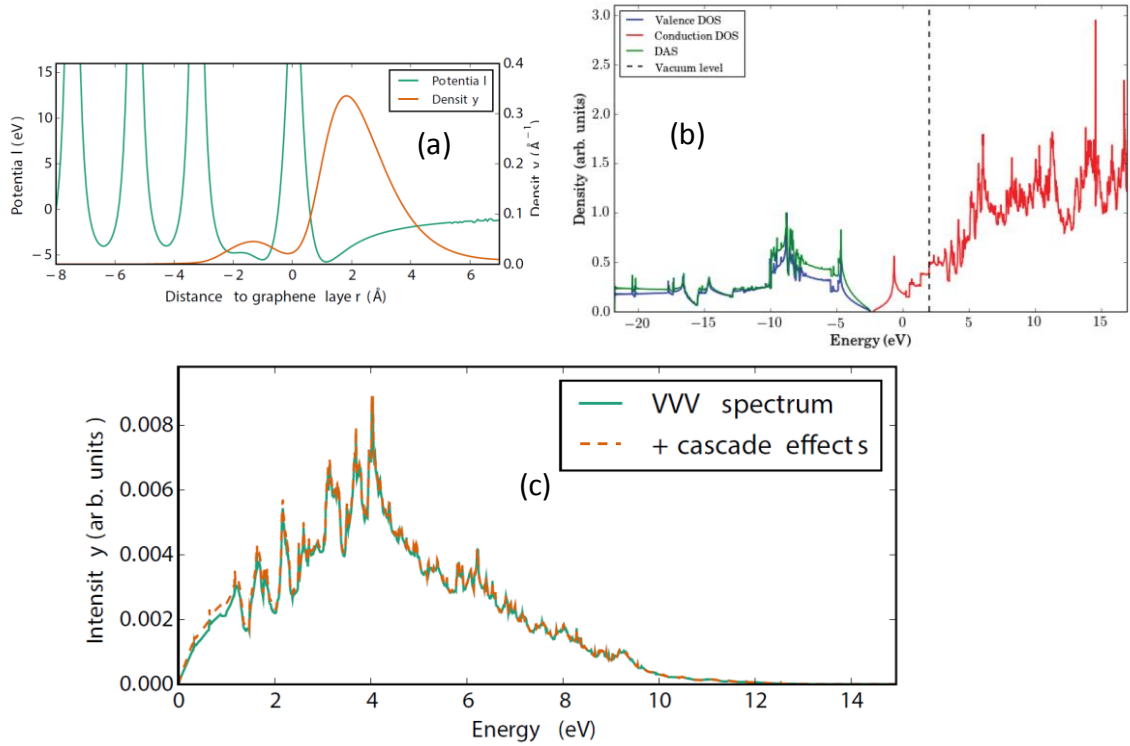

Fig. 3 (a) Plot of the positron potential (green) and the groudstate (orange) for a single graphene layer on Cu (111) substrate. Our result shows that a single graphene layer on Cu (111) supports a positron surface state bound in its image potential well. (b) Density of the electronic valence (blue) and conduction (red) states. The vacuum level is indicated by the dashed line. The green curve shows the distribution of holes that initiate the VVV Auger process (density of annihilating states). (c) Auger electron spectrum as calculated by our extension of the formalism by Hagstrum [12] for a free standing graphene layer (green). In orange, we show the result in which cascade effects are taken into account, i.e. each hole created in the Auger process initiates an additional Auger transition. The cascade effect has only a small quantitative influence on the spectrum..

.

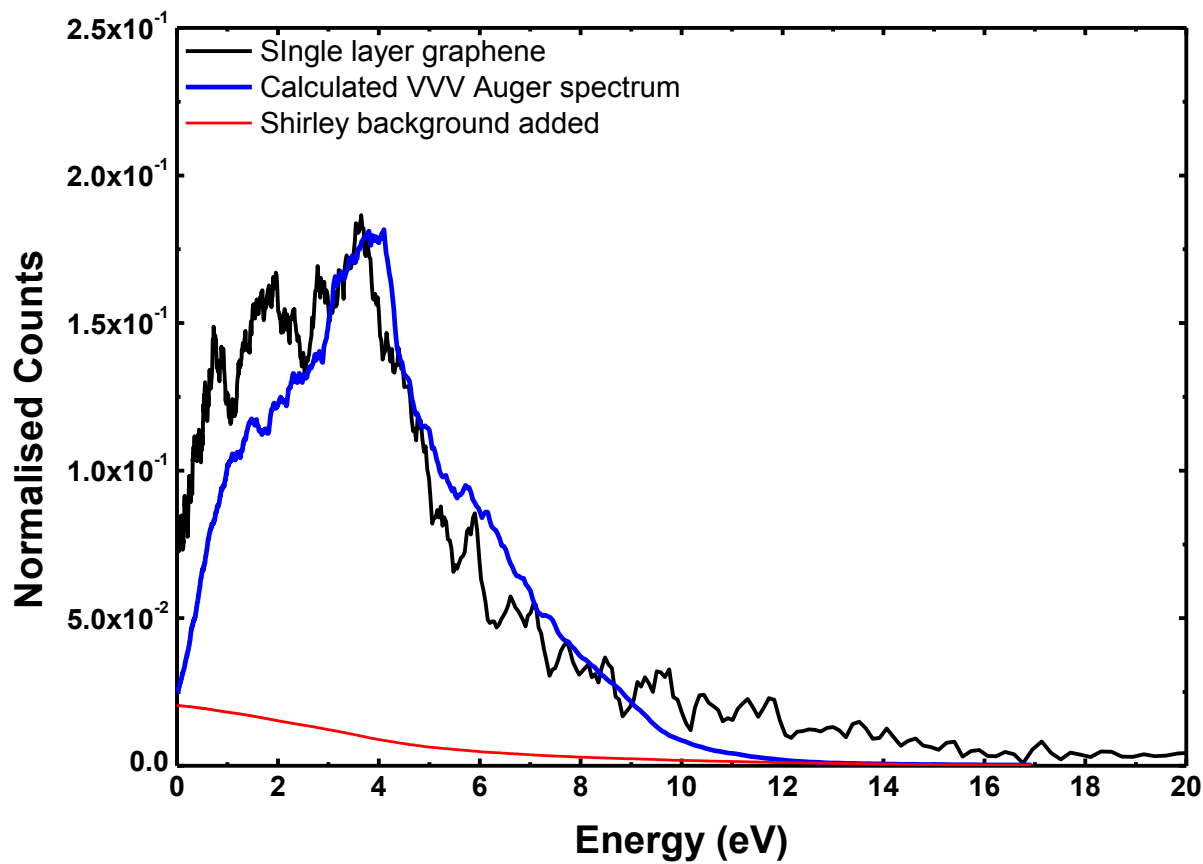

Fig.4 A comparison of the measured and calculated energy distribution of electrons emitted via an annihilation induced VVV Auger transition from a single layer of graphene on Cu. The inelastic tail from high energy Auger peaks has been subtracted from the measured distribution. The calculated Auger spectrum shown in 3 (c) has been broadened using a simulated instrumental response function of the TOF spectrometer system. The best agreement was found after adding a small contribution from electrons assumed to be inelastically scattered as they exit the surface modelled using a Shirley function shown in red.

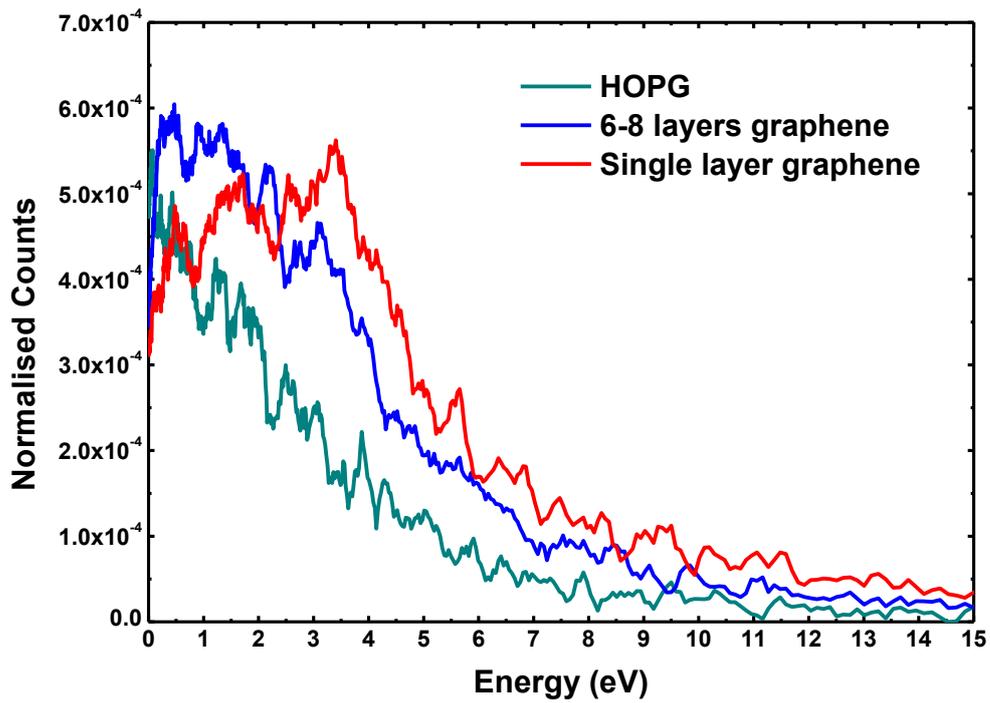

Fig.5 A comparison of the distribution of positron annihilation induced low energy electrons emitted from single layer graphene to the distribution measured from HOPG and multi-layer graphene (6-8) layers. The shift of the spectrum to lower intensity is consistent with an expected downward shift in the energy distribution of electrons emitted due to annihilation induced VVV Auger processes as a result of reduced substrate screening of final-state hole-hole interactions [18].